\documentclass[twocolumn,aps]{revtex4}
\usepackage{graphicx}

\begin{document}
\title{Illustration of the Jarzynski nonequilibrium work relation for an ideal gas}
\author{Rhonald C. Lua}
\affiliation{Department of Physics, University of Minnesota \\
116 Church Street SE, Minneapolis, MN 55455}

\begin{abstract}
The Jarzynski relation is a recently discovered result relating
the average exponential of the work done under nonequilibrium conditions
to an equilibrium free energy difference. We illustrate this remarkable relation
by considering the expansion and compression of a classical
ideal gas particle inside a cavity with a moving piston in one dimension.
Using a simple relationship between the distributions of the nonequilibrium work done
in the two (forward and reverse) processes, we verify the Jarzynski relation and
discuss the practical realization of the Jarzynski equality for the model.
\end{abstract}

\maketitle

\section{Introduction}
Most relations in thermodynamics are valid under equilibrium or
near-equilibrium conditions. For example, consider a system in contact
with a heat bath at temperature $T$.
If under some (infinitesimal) process
an energy $Q$ is exchanged with
the bath, the first law of thermodynamics tells us that
\begin{equation}
Q=\Delta U + W \ ,\label{firstlaw}
\end{equation}
where $\Delta U$ is the change in the internal energy of the system
and $W$ is the work done by the system.
The statement (\ref{firstlaw}) is of course an expression of the conservation
of energy, which is valid not just in equilibrium or whether the process
is reversible or irreversible.
If we assume that the process takes place
very slowly (quasistatic, reversible), the second law of thermodynamics further
relates the
entropy change $\Delta S$ to the energy $Q$ via the relation $Q=T\Delta S$.
With this replacement, the relationship (\ref{firstlaw}) may be recast as
\begin{equation}
-W=\Delta U - T\Delta S=\Delta F
\label{WdF}
\end{equation}
relating the work done on the system $-W$ to the change in the
Helmholtz free energy $\Delta F$.

If the process is not quasistatic, we have from the second law of thermodynamics,
$\left< Q \right>\le T\Delta S$, and the equality (\ref{WdF}) converts into an inequality
\begin{equation}
-\left< W \right> \ge \Delta F
\label{WdF2}
\end{equation}
where the $<\cdots>$ denotes an average over an ensemble of measurements
of $W$ (or $Q$), where each measurement is taken as the system
undergoes the same variation in external parameters starting from an equilibrium state.
(For the system of a gas under a piston and the process consisting of movement of the piston, this means that for each measurement the piston
moves along the same trajectory, with the gas in thermal equilibrium
at the start of every trajectory.
The external parameter in this case is the volume of the enclosure, which is the
same at the start of every trajectory and the same at the end of every trajectory.)

Many macroscopic physical quantities measured in experiments where the system
is driven away from equilibrium satisfy inequalities such as (\ref{WdF2}).
However, in 1997, Jarzynski \cite{Jar1} announced to the scientific community the following 
\textit{equality} that now bears his name
\begin{equation}
\left< e^{W/k_BT} \right>=e^{-\Delta F/k_B T}
\label{Jar}
\end{equation}
where the angle brackets have the same meaning as in the previous paragraph.
This remarkable formula tells us that we can extract equilibrium information ($\Delta F$)
from the ensemble of \textit{nonequilibrium} (finite-time) measurements \cite{Jar1}.
By application of the mathematical identity $\left < e^{x} \right> \ge e^{\left< x \right>}$
the Jarzynski relation (\ref{Jar}) yields the inequality in (\ref{WdF2}).

For our purposes, the Jarzynski relation is valid when
at the start of every measurement the system is at thermal
equilibrium with a heat bath, and that during the finite-time switching of the external parameters, no energy is exchanged with the heat bath (i.e. the process is adiabatic).
The free energies $F_1$ and $F_0$ that comprise the difference $\Delta F=F_1-F_0$
refer to states of the system in equilibrium at the external parameters
corresponding to the start and end of the switching process.
More detailed and general conditions for the validity of (\ref{Jar}) can
be found in \cite{Jar1,Jar2,Crooks1,Crooks2,FrenkelSmit}.

The utility of the Jarzynski relation lies in its potential to give a good estimate
of the free energy change, $\Delta F$, given a finite number of measurements
of the work within a finite amount of time.
(e.g. compared with the use of (\ref{WdF2}))
In fact, the Jarzynski relation
has been used to measure free energy differences
in experiments
involving single molecule systems such as the pulling of RNA or protein \cite{Jar3,Hummer1,Liphardt,PhysToday}.

\begin{figure}
\centerline{\scalebox{0.4}{\includegraphics{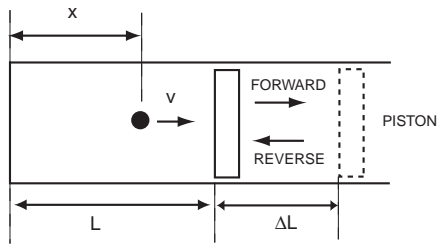}}}
\caption
{Definition of the forward (expansion) and reverse (contraction) processes in the piston-cavity
system.} \label{fig:forwardreversepiston}
\end{figure}

Nevertheless, it should be pointed out that
there exists considerable recent debate in the literature concerning when precisely
the Jarzynski
relation is valid and whether it is applicable in practice, even for an ideal gas
\cite{LuaGrosberg,Gross1,Jar4,Gross2,Sung,Jepsen}.
In reference \cite{LuaGrosberg}, Lua and Grosberg
discussed the practical applicability of the Jarzynski relation in the context of
an ideal gas particle expanding under a piston and moving in one dimension.
(See figure \ref{fig:forwardreversepiston}. The Jarzynski equality for $N$ independent
particles is given by the Jarzynski equality for $1$ particle raised to the power $N$.)
In that work, the Jarzynski relation was verified with an explicit calculation.
The work distribution function was also derived and given explicitly.
Finally, the limit of a fast moving piston was examined, leading to the conclusion
that it might be difficult to realize the Jarzynski equality in practice for the
case of an expansion with the fast moving piston,
as was also pointed out in \cite{Gross2}.

Our goal in this article is to illustrate the Jarzynski relation and gain some
insight into how it works by considering the same ideal gas model discussed
in \cite{LuaGrosberg}.
Our results and conclusions for the case of an expanding cavity
are the same as in that work.
However, in the present work
we also examine the `reverse' case in which the piston compresses the cavity.
Furthermore, we verify the Jarzynski relation in an indirect but simpler fashion.

The plan of the rest of the article is as follows.
In section \ref{sec:prob}, we give explicit expressions for the
probability distribution functions of the work
done by the gas particle (or piston) for the case when the piston expands
the cavity (arbitrarily called the `forward' process, with distribution denoted
by $P_F(W)$, where the argument $W$ is the magnitude of the work done by the gas)
and for the case when the piston compresses the cavity
(called the `reverse' process, with distribution denoted by $P_R(W)$, where $W$ is
the work done by the piston).
These probability distributions appear when calculating the average
defined above, e.g. $\left< e^{W/k_BT} \right>_F=\int_0^\infty P_F(W)e^{W/k_BT}dW $.
We assume that the piston moves with the same
constant speed during the expansion and compression.
In an attempt to make the main presentation short and simple,
we shall also relegate cumbersome derivations to the Appendix.
In section \ref{sec:verify}, we show that the forward and reverse distributions,
$P_F(W)$ and $P_R(W)$, satisfy a simple relation, which we then use to show
that the Jarzynski relation is valid for our model. In section \ref{sec:eff},
we discuss the practical evaluation of the free energy difference from the
exponential average in the Jarzynski relation
for the forward and reverse processes.

\section{Probability distribution of the work}\label{sec:prob}

Referring to figure \ref{fig:forwardreversepiston},
the initial state of the particle $(x,v)$ consists of a position $x$ 
distributed with uniform probability within the initial length of the cavity, and
a velocity $v$ drawn from a Maxwell-Boltzmann distribution (i.e. a Gaussian distribution,
 $e^{-v^2/2}$).
To simplify the writing and without loss of generality, we use units such that
$k_BT=1$, the mass of the particle $m=1$, and that the piston displaces by
an amount $\Delta L$ during time $\tau=1$.

For the forward process in which the cavity expands from `volume' $L$ to a final volume
of $L+\Delta L$, the prescription for evaluating the distribution is
\begin{equation}
P_F(W)=\frac{1}{\sqrt{2\pi}L}\int_0^L dx\int_{-\infty}^{\infty}dv
e^{-v^2/2}\delta \left( W-w_{\tau}(x,v) \right) \ ,
\label{presforward}
\end{equation}
where $w_\tau(x,v)$ is the work done by the particle given the initial coordinate $(x,v)$.
The $\delta$ function picks out the initial states which lead to an amount of
work $W$.
The calculation is presented in detail in Appendix \ref{sec:derivation}. The final result is
\begin{equation}
P_F(W) = \delta(W) P_{0} +\frac{e^{-\frac{1}{2}\left(n\Delta L+\frac{W}{2n\Delta L}\right)^2}}{\sqrt{2\pi}n\Delta L}f(W) \ . \label{distforward}
\end{equation}
Here, $P_{0}$ is the probability to obtain vanishing work because
the particle is unable to chase the piston or hit it even once,
\begin{equation} P_{0} = \frac{1}{\sqrt{2\pi}L}\int_0^Ldx\int_{-(L+\Delta L)}^{(L+\Delta L)}dv e^{-(v-x)^2/2} \ .
\end{equation}
The integer $n$ (which is the number of bounces by the particle against the piston) is
obtained in Appendix \ref{sec:derivation} and is given by the formula
\begin{equation}
n = \left[
\left.\left(1+\sqrt{1+\frac{2 W}{\Delta L(2L+\Delta L)}} \right) \right/ 2
\right] \ , \label{eq:expression_for_n}
\end{equation}
where $\left[ \ldots \right]$ means integer part of $\ldots$.  For
example, simple algebra indicates that as long as $W < 4 \Delta L(2 L +\Delta L)$, we have just one collision, $n=1$.  For the values of work
$W$ in the next interval, $4 \Delta L(2 L +  \Delta L) < W < 12 \Delta L(2 L +\Delta L) $,
 we have $n=2$, etc.

Finally, the function $f(W)$, which we call the overlap factor, varies between
$0$ and $1$ and is illustrated
in figure \ref{fig:overlapplot}. An explicit expression is given
by (\ref{fexplicit}) in Appendix \ref{sec:derivation}.

Similarly, for the reverse process
in which the cavity contracts from `volume' $L+\Delta L$ to a final volume
of $L$, the prescription for evaluating the distribution is
\begin{eqnarray}
P_R(W)&=&\frac{1}{\sqrt{2\pi}(L+\Delta L)}\int_0^{L+\Delta L} dx \times \nonumber\\
&&\times \int_{-\infty}^{\infty}dv e^{-v^2/2} \delta \left( W-w'_{\tau}(x,v) \right) \ ,
\label{presreverse}
\end{eqnarray}
where $w'_\tau(x,v)$ is the work done by the piston given the initial coordinate of
the particle $(x,v)$.
The final result is
\begin{equation}
P_R(W) = \delta(W) P'_{0} +\frac{e^{-\frac{1}{2}\left(n\Delta L-\frac{W}{2n\Delta L}\right)^2}}{\sqrt{2\pi}n\Delta L}\frac{L f(W)}{(L+\Delta L)} \ . \label{distreverse}
\end{equation}
Here, $P'_{0}$ is given by
\begin{equation} P'_{0} = \frac{1}{\sqrt{2\pi}(L+\Delta L)}\int_0^{L+
\Delta L}dx\int_{-L}^{L}dv e^{-(v-x)^2/2} \ ,
\end{equation}
The integer $n$ and the function $f(W)$ are the same as in the forward case.

Thus, the probability distributions (\ref{distforward}) and (\ref{distreverse})
consist of a $\delta$ function peak at $W=0$ and a tail or hump at positive $W$
(see figure \ref{fig:forwardreversecurves} for the case of a fast moving piston).
The forward distribution is examined in Appendix \ref{sec:fast} in the limit of a fast moving piston
and in Appendix \ref{sec:slow} in the limit of a slow moving piston.

\section{Verification of the Jarzynski relation}\label{sec:verify}

We now show that our model satisfies the Jarzynski relation (\ref{Jar}).
Taking the ratio of expressions (\ref{distforward}) and (\ref{distreverse}),
we obtain
\begin{equation}
\frac{P_F(W)}{P_R(W)}=e^{-W-\Delta F}  \ , \label{fluct}
\end{equation}
where $\Delta F=-\ln{\left(\frac{L+\Delta L}{L}\right)}$ is the difference between the
free energies at volumes $L+\Delta L$ and $L$, in equilibrium at the same temperature.
(Relation (\ref{fluct}) is reminiscent of another remarkable result called
the fluctuation theorem or Crooks relation,
which relates the distribution of entropy productions of a driven system
that is initially in equilibrium to the entropy production of the same system
driven in reverse \cite{Crooks1,Crooks2,PhysToday,Wang,FrenkelSmit}. The simple proof below was also
inspired by the first two of these references.).

The exponential average of the work is readily evaluated from (\ref{fluct})
and the normalization of the probability distributions. For the forward process,
\begin{eqnarray}
\left< e^W \right>_F&=&\int_0^\infty P_F(W)e^WdW \nonumber\\
&=&e^{-\Delta F}\int_0^\infty P_R(W) dW \nonumber\\
&=&e^{-\Delta F}
\end{eqnarray}
Similarly, for the reverse process
\begin{eqnarray}
\left< e^{-W} \right>_R&=&\int_0^\infty P_R(W)e^{-W}dW \nonumber\\
&=&e^{\Delta F}\int_0^\infty P_F(W) dW \nonumber\\
&=&e^{\Delta F}
\end{eqnarray}

The Jarzynski relation is also verified explicitly in appendices \ref{sec:fast} and \ref{sec:slow} in the opposite
limits of a fast moving piston and a slow moving piston.

\section{Realizing the Jarzynski equality in practice}\label{sec:eff}
In this section we examine how well the average of the left-hand-side of
the Jarzynski relation (\ref{Jar}) can reproduce the equality with the right-hand-side, based on a finite-number of measurements of the work. This issue is of immense
practical interest in experiments involving the manipulation of nanoscale objects
and biological molecules.

From the definition of the exponential average, $\left< e^{W} \right>_F=\int_0^\infty P_F(W)e^{W}dW $, one can already see the potential problem. In an expansion, when
the piston moves very fast, only the very fast (and very rare) particles
can hit the piston to produce non-zero work. But these fast particles can give a
significant contribution to the average because of the factor $e^W$. Many
measurements or repetitions of the expansion are necessary to
sample an adequate number of particles that move fast enough in order to
assess this contribution accurately.

To be more precise, and to see a clear difference between the forward
and reverse processes, we consider the case when the piston moves very fast, such
that a typical particle moves very little
during the expansion or compression of the cavity. This is equivalent to
the conditions $L\gg c\tau$ and $\Delta L \gg c\tau$, where
$c=\sqrt{k_B T/m}$ is roughly the speed of sound in the gas. In our simplified units, this translates
to $L,\Delta L \gg 1$.

First, let us rewrite (\ref{fluct}) as
\begin{equation}
P_F(W)e^W=P_R(W)e^{-\Delta F}
\end{equation}
we see that the contribution of the expanding or forward work measurements to the average, $e^{-\Delta F}$, is given
by the reverse work distribution, $P_R(W)$. This indicates that measurements of $P_F(W)$ can give
an accurate result for the average if $P_F(W)$ is significant at values of $W$ where $P_R(W)$ is significant; that is, if
there is a reasonable overlap between the forward and reverse work distributions
\cite{FrenkelSmit}.

Now let us determine the probability of obtaining zero work for the forward and reverse
processes. For the forward process, the probability $P_F(0)$ is clearly close to $1$.
From the relation (\ref{fluct}), the probability of obtaining zero work for
the reverse process is given by,
\begin{equation}
P_R(0) \approx \frac{L}{L+\Delta L}
\end{equation}
Clearly, in the limit of a fast moving piston, this probability is given by
the fraction of the cavity volume that is not swept or visited by the piston.

\begin{figure}
\centerline{\scalebox{0.4}{\includegraphics{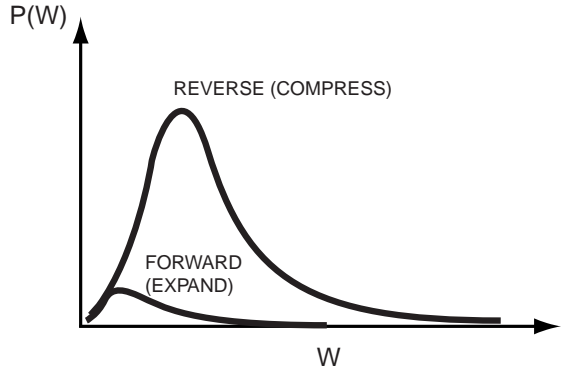}}}
\caption
{Probability distribution for the non-zero work values in the forward
and reverse processes, when the piston moves fast.} \label{fig:forwardreversecurves}
\end{figure}

As to the probability of obtaining non-zero work, in the forward process
this clearly vanishes (the area under the lower curve in figure \ref{fig:forwardreversecurves}) in the limit of a fast moving piston.
In contrast, the probability of obtaining non-zero work in the reverse process is finite
(the area under the taller curve in figure \ref{fig:forwardreversecurves}).
This probability is given by the fraction of the cavity volume that is swept by the piston,
$\Delta L/(L+\Delta L)$.

Now we can address the issue presented at the start of this section.
In the forward process, rare events (non-zero work values) contribute a finite
fraction to the average, given by the area under the `reverse' curve of figure \ref{fig:forwardreversecurves}. This means that many measurements are necessary
(given roughly by the inverse probability of obtaining non-zero work values (\ref{problargevp}))
in order to faithfully produce the equality in the Jarzynski relation (\ref{Jar}).
In contrast, for the reverse process, the dominant contribution is given by the zero
work values, which have finite probability. Therefore a smaller number of measurements in the reverse process
can reproduce the equality in (\ref{Jar}) quite well, as was indeed revealed in computer
simulations.

\section{Conclusion}
We obtained explicit expressions for the distribution of the work done
in the forward and reverse processes,
$P_F(W)$ and $P_R(W)$, for a simple ideal gas model.
By noting a simple relationship between the two distributions (equation (\ref{fluct}))
we illustrated the validity of the Jarzynski relation (equation (\ref{Jar})) for the model.
We also conclude that in the forward or expanding case,
the evaluation of the average exponential of the nonequilibrium work
from a finite number of trials or measurements can give a poor result
for the free energy difference when the piston moves sufficiently fast
(indicating that repeated, rapid nonequilibrium measurements may not be
automatically advantageous \cite{LuaGrosberg}).
This is in sharp contrast to the average that can be obtained from the
complementary or reverse process. In general an optimal result
may be obtained by a suitable combination of the measurements in the forward
and reverse processes \cite{Shirts}.

\section*{Acknowledgments}
I thank my research adviser, Alexander Yu. Grosberg, for introducing me to
the Jarzynski relation and its application to the ideal gas model discussed here,
which pointed to the role of the far tails of the Maxwell distribution in resolving
a paradox involving a very fast moving piston. I also appreciate his comments
on this manuscript.

\appendix
\section{Calculation of $P_F(W)$}\label{sec:derivation}
In this section, we first derive an expression for $P_F(W)$,
based on the work in \cite{LuaGrosberg}. Again, we take reduced units
such that $k_BT=1$, the particle mass $m=1$, and
the piston moves with speed $\Delta L/\tau$ during a time interval $\tau=1$.

Let us first assume a positive initial velocity (refer to figure \ref{fig:forwardreversepiston}),
in which the
particle can strike the piston first before hitting the left end
of the cavity. The time taken for the first collision with the
piston is $t_1=\frac{L-x}{v-\Delta L}$. After the collision, the
velocity of the particle relative to the piston gets reversed and
the speed of the particle gets diminished to $v-2\Delta L$ (assuming $v>2\Delta L$).
The time taken for the second collision with the piston
is given by $t_2=\frac{3L-x}{v-3\Delta L}$.
In general, for the
$n^{th}$ collision
\begin{equation}
t^{+}_{n}=\frac{(2n-1)L-x}{v-(2n-1)\Delta L}
\end{equation}
Similarly, for a particle with a negative initial velocity,
\begin{equation}
t^{-}_{n}=\frac{(2n-1)L+x}{v-(2n-1)\Delta L}
\end{equation}

These relations can be inverted to give conditions that should be
satisfied by the speed of the particle in order to result in
exactly $n$ collisions with the piston within a time interval
$\tau = 1$. For positive initial velocities,
\begin{equation}
(2n-1)(L+\Delta L)-x<|v|<(2n+1)(L+\Delta L)-x\label{eq:ineq1}
\end{equation}
For negative initial velocities,
\begin{equation}
(2n-1)(L+\Delta L)+x<|v|<(2n+1)(L+\Delta L)+x \label{eq:ineq2}
\end{equation}

The work done by the piston on the particle after one collision is
the change in momentum of the particle times the velocity of the
piston,
\begin{equation}
-w_1=(-(v-2\Delta L)-v)\Delta L=-2(v-\Delta L)\Delta L
\end{equation}
In general, the work done after $n$ collisions is
\begin{equation}
-w_n=-2v\Delta L n+2{\Delta L}^2n^2
\end{equation}
Note that the work done can also be calculated from the change in kinetic energy after $n$ collisions,
\begin{equation}
-w_n=\frac{1}{2}(v-2n\Delta L)^2-v^2/2=-2n\Delta L v+2n^2{\Delta L}^2
\label{eq:work_in_n}
\end{equation}
The work done by the particle on the piston is positive for an
expanding volume.

The inequalities (\ref{eq:ineq1}) and (\ref{eq:ineq2})
lead to the following partition of
the integral in (\ref{presforward}),
\begin{widetext}
\begin{eqnarray}
P_F(W)&=& \frac{1}{\sqrt{2\pi}L} \int_0^Ldx\sum_{n=1}^{\infty} \int_{(2n-1)(L+{\Delta L})-x}^{(2n+1)(L+{\Delta L})-x}dv e^{-v^2/2} \times \delta \left( W- (2v{\Delta L}n-2{\Delta L}^2n^2) \right) + \nonumber\\
&&\frac{1}{\sqrt{2\pi}L}\int_0^Ldx\sum_{n=1}^{\infty} \int_{(2n-1)(L+{\Delta L})+x}^{(2n+1)(L+{\Delta L})+x}dv e^{-v^2/2} \times \delta \left( W-(2v{\Delta L}n-2{\Delta L}^2n^2) \right) + \nonumber\\
&&\frac{1}{\sqrt{2\pi}L}\int_0^Ldx\int_{-(L+{\Delta L})-x}^{(L+{\Delta L})-x}dv
e^{-v^2/2}\delta(W-0)\nonumber
\end{eqnarray}
\end{widetext}
Call the first term $I_1$ and the second term $I_2$ (the third
term is `trivial'). Performing a change of variable to remove $x$
from the limits, integrating over $x$ and taking advantage of the
delta function results in
\begin{eqnarray}
I_1&=&\sum_{n=1}^{\infty}\frac{e^{-\frac{1}{2}\left(n{\Delta L}+\frac{W}{2n{\Delta L}}\right)^2}}{\sqrt{2\pi}n{\Delta L}}\times \nonumber\\
&&\times \frac{1}{2L}\times \left\{\mbox{overlap between}\right.\nonumber\\
&&\left.\left[n{\Delta L}+\frac{W}{2n{\Delta L}},n{\Delta L}+\frac{W}{2n{\Delta L}}+L\right]\right. \nonumber\\
&&\left.\mbox{and}\left[(2n-1)(L+{\Delta L}),(2n+1)(L+{\Delta L})\right]
\right\}\nonumber
\end{eqnarray}
and similarly for $I_2$,
\begin{eqnarray}
I_2&=&\sum_{n=1}^{\infty}\frac{e^{-\frac{1}{2}\left(n{\Delta L}+\frac{W}{2n{\Delta L}}\right)^2}}{\sqrt{2\pi}n{\Delta L}}\times \nonumber\\
&&\times \frac{1}{2L} \times \left\{\mbox{overlap between}\right.\nonumber\\
&&\left.\left[n{\Delta L}+\frac{W}{2n{\Delta L}}-L,n{\Delta L}+\frac{W}{2n{\Delta L}}\right]\right. \nonumber\\
&&\left.\mbox{and}\left[(2n-1)(L+{\Delta L}),(2n+1)(L+{\Delta L})\right]
\right\}\nonumber
\end{eqnarray}
(For two intervals $[a,b]$ and $[c,d]$ where $a<b$ and $c<d$,
the expression $\left\{\mbox{overlap between} [a,b] \mbox{and} [c,d] \right\}$
equals $0$ when $b<c$, equals $b-c$ when $a<c<b<d$, etc.)
$I_1$ and $I_2$ can be combined as follows,
\begin{eqnarray}
I_1+I_2&=&\sum_{n=1}^{\infty}\frac{e^{-\frac{1}{2}\left(n{\Delta L}+\frac{W}{2n{\Delta L}}\right)^2}}{\sqrt{2\pi}n{\Delta L}}\times \nonumber\\
&&\times \frac{1}{2L} \times \left\{\mbox{overlap between}\right.\nonumber\\
&&\left.\left[n{\Delta L}+\frac{W}{2n{\Delta L}}-L,n{\Delta L}+\frac{W}{2n{\Delta L}}+L\right]\right. \nonumber\\
&&\left.\mbox{and}\left[(2n-1)(L+{\Delta L}),(2n+1)(L+{\Delta L})\right] \right\}\nonumber\\
&=&\sum_{n=1}^{\infty}\frac{e^{-\frac{1}{2}\left(n{\Delta L}+\frac{W}{2n{\Delta L}}\right)^2}}{\sqrt{2\pi}n{\Delta L}}\times
f(n,W) \nonumber
\end{eqnarray}
where the overlap factor $f(n,W)$ satisfies $0 \leq f \leq 1$, since
the range of the smaller interval is at most $2L$. $f$ is also
zero for negative $W$, or positive work values $-W$ done by the piston.
The conditions that must be satisfied by $W$ in order for the
overlap associated with integer $n$ to occur are
\begin{widetext}
\begin{eqnarray}
2n{\Delta L}\left(2(n-1)L+(n-1){\Delta L}\right)<W<2n{\Delta L}\left(2(n+1)L+(n+1){\Delta L}\right)
\label{wint}
\end{eqnarray}
\end{widetext}
Notice that the left boundary of the interval is a function of
$n(n-1)$, while the right interval is a function of $n(n+1)$.
Therefore the right boundary can be transformed into the
left-boundary by making the replacement $n\rightarrow n-1$. This
implies that the intervals (\ref{wint}) are contiguous and
nonoverlapping and that at most one term in the summation in
$P_F(W)$ survives. One can solve for the integer $n$ by taking the
integer part (or floor function) of a solution to a quadratic
equation,
\begin{equation}
W > 2n{\Delta L}\left(2(n+1)L+(n+1){\Delta L} \right)
\end{equation}
which results in formula (\ref{eq:expression_for_n}) in the main
text.

Dropping the argument $n$, an explicit expression for $f(W)$ is
\begin{widetext}
\begin{equation}
f(W) = \left\{
\begin{array}{lcr}
-(n-1) \left(\frac{{\Delta L}}{2L}+1 \right) + \frac{W}{4n{\Delta L}L} & {\rm when} &
(n-1)({\Delta L}+2L) < \frac{W}{2n{\Delta L}} \leq (n-1)({\Delta L}+2L)+2L\\
1 & {\rm when}  &
(n-1)({\Delta L}+2L)+2L < \frac{W}{2n{\Delta L}} \leq (n-1)({\Delta L}+2L)+2L+2{\Delta L}\\ (n+1)\left(\frac{{\Delta L}}{2L}+1 \right)-\frac{W}{4n{\Delta L}L} & {\rm when} & (n-1)({\Delta L}+2L)+2L+2{\Delta L} < \frac{W}{2n{\Delta L}} \leq (n+1)({\Delta L}+2L)
\end{array} \right.
\label{fexplicit}
\end{equation}
\end{widetext}

\section{Obtaining $P_R(W)$ from $P_F(W)$}
To obtain $P_R(W)$ from $P_F(W)$, make the following series of replacements
$W \rightarrow -W$, $\Delta L \rightarrow -\Delta L$, $L \rightarrow L+\Delta L$.
The result is
\begin{widetext}
\begin{eqnarray}
P_R(W)&=&P'_0\delta(W)+\frac{e^{-\frac{1}{2}\left(-n{\Delta L}+\frac{W}{2n{\Delta L}}\right)^2}}{\sqrt{2\pi}n{\Delta L}}\times \frac{1}{2(L+\Delta L)}\times \nonumber\\
&&\times \left\{\mbox{overlap between}
\left[-n{\Delta L}+\frac{W}{2n{\Delta L}}-(L+{\Delta L}),-n{\Delta L}+\frac{W}{2n{\Delta L}}+L+{\Delta L}\right]\mbox{and}\left[(2n-1)(L),(2n+1)(L)\right] \right\}\nonumber
\end{eqnarray}
\end{widetext}
where
\begin{equation}
P'_0 = \frac{1}{\sqrt{2\pi}(L+{\Delta L})}\int_0^{L+{\Delta L}}dx\int_{-L}^{L}dv e^{-(v-x)^2/2}
\end{equation}
and
\begin{equation} n = \left[
\left.\left(1+\sqrt{1+\frac{2 W}{{\Delta L}(2L+{\Delta L})}} \right) \right/ 2
\right]
\end{equation}
Notice that $n$ on the forward and reverse cases are identical.

The double integral in the $P_0$ term in the forward process is identical to
the double integral in the $P'_0$ term due to reflection symmetry of the integrand,
which means
\begin{equation}
\frac{P_0}{P'_0}=\frac{L+\Delta L}{L}
\end{equation}

Furthermore the overlaps of the intervals in the forward and reverse processes are identical.
To see this, shift each interval in the overlap
expression for the reverse case by $(2n-1){\Delta L}$ (this does not change the value of the overlap)
\begin{widetext}
\begin{eqnarray}
\left\{\mbox{overlap between}\left[-n{\Delta L}+\frac{W}{2n{\Delta L}}-(L+{\Delta L}),-n{\Delta L}+\frac{W}{2n{\Delta L}}+L+{\Delta L}\right]
\mbox{and}\left[(2n-1)(L),(2n+1)(L)\right] \right\} = \nonumber\\
= \left\{\mbox{overlap between}\left[n{\Delta L}+\frac{W}{2n{\Delta L}}-L-2{\Delta L},n{\Delta L}+\frac{W}{2n{\Delta L}}+L\right] \mbox{and} \left[(2n-1)(L+{\Delta L}),(2n+1)(L+{\Delta L})-2{\Delta L}\right] \right\}\nonumber\\
\end{eqnarray}
\end{widetext}
Comparing this expression for the overlap in the reverse case with that of the 
forward case, one sees that they are identical apart from the shift of $-2{\Delta L}$
in the left boundary of the first interval and a shift of $-2{\Delta L}$ on the right boundary
of the second interval. Because the difference of the lengths of the two intervals is
$2{\Delta L}$, the overlaps are identical.

\section{Limit of a fast moving piston}\label{sec:fast}
In this section, we obtain the form of the forward distribution (\ref{distforward}) when the
piston takes a very short time $\tau$ to undergo the displacement $\Delta L$,
such that a typical particle moves very little during the expansion.
That is, we assume that $\Delta L \gg c\tau$ and $L \gg c\tau$, where $c=\sqrt{k_BT/m}$.
The formulas in this section are similar to that in \cite{LuaGrosberg}.

In this limit, the distribution is dominated by a single bounce, $n=1$,
and values $W<4(\Delta L)L$, giving
\begin{equation}
P(W) \simeq \delta(W) P_{0} +
\frac{e^{-\frac{1}{2}\left({\Delta L}+\frac{W}{2{\Delta L}}\right)^2}}{\sqrt{2\pi}{\Delta L}}
\frac{W}{4 ({\Delta L})L} \ . \label{eq:distslow}
\end{equation}
This form of the distribution suffices to produce the Jarzynski equality:
\begin{eqnarray}
\langle e^W \rangle &=&\int_0^{\infty} P(W)e^W dW\nonumber\\
&=&
P_{0}+\frac{1}{\sqrt{2\pi}{\Delta L}}\int_0^{\infty}e^{-\frac{1}{8{\Delta L}^2}(W-2{\Delta L}^2)^2}\times\nonumber\\
&&\times\frac{W}{4{\Delta L}L}dW\nonumber\\
&=&\int_0^{L}\frac{dx}{L}\frac{1}{\sqrt{2\pi}}\int_{-\infty}^{\infty}dv e^{-\frac{1}{2}(v-x)^2}+\nonumber\\
&&\frac{1}{\sqrt{2\pi}{\Delta L}}\int_{-\infty}^{\infty}e^{-\frac{1}{8{\Delta L}^2}(W-2{\Delta L}^2)^2}\times\nonumber\\
&& \times \frac{2{\Delta L}^2}{4({\Delta L})L}dW\nonumber\\
&=&1+\frac{{\Delta L}}{L}=e^{-\Delta F} \ . \nonumber
\end{eqnarray}
Where $\Delta F=-\ln{(1+\Delta L/L)}$. Here, we made approximations in both the $P_0$ term, by extending the
integral limits to $(-\infty,\infty)$, and in the tail term, by
setting the integral lower limit to $-\infty$.

Let us also calculate the probability of obtaining non-zero work
values.
Using expression (\ref{eq:distslow}), we have
\begin{eqnarray}
P_{W>0} &=& \int_{0^+}^\infty P(W)dW\nonumber\\
&=&\frac{e^{-{\Delta L}^2/2}}{4\sqrt{2\pi}L({\Delta L})^2}\int_0^\infty W\times\nonumber\\
&& \times e^{-\frac{1}{2}W-\frac{1}{8{\Delta L}^2}W^2}dW
\end{eqnarray}
For large ${\Delta L}$, we neglect the term $W^2/(8{\Delta L}^2)$ in the exponent upon integration, yielding
\begin{eqnarray}
P_{W>0} &\simeq& \frac{1}{\sqrt{2\pi}L({\Delta L})^2}e^{-{\Delta L}^2 / 2} \label{problargevp}
\end{eqnarray}
This quantity represents the area under the smaller curve in figure \ref{fig:forwardreversecurves}.

\section{Limit of a slow moving piston}\label{sec:slow}
In this section, we obtain the form of the forward distribution (\ref{distforward}) when the
piston takes a very long time $\tau$ to undergo the displacement $\Delta L$,
such that a typical particle bounces off the piston many times.
This means that $\Delta L \ll c\tau$ and $L \ll c\tau$, where $c=\sqrt{k_BT/m}$.

\begin{figure}
\centerline{\scalebox{0.4}{\includegraphics{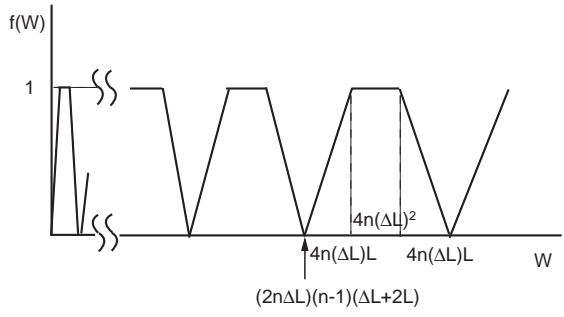}}}
\caption[Plot of the overlap factor $f(W)$.]
{The structure of the overlap factor $f(W)$ that modulates the exponential in the
distribution function. This factor becomes a rapidly oscillating function
in the limit of small piston velocities (i.e. a quasistatic process).} \label{fig:overlapplot}
\end{figure}

Defining $\Delta L/L=\alpha$,
we have $P_0\simeq 0$ and $n$ is very large with
$n\Delta L\simeq \sqrt{\frac{W}{2\alpha(2+\alpha)}}\alpha$.
Thus,
\begin{equation}
P(W) \simeq
\frac{\sqrt{\alpha(2+\alpha)/\pi}}{\alpha\sqrt{W}}\exp{\left(-\frac{(1+\alpha)^2}{\alpha(2+\alpha)}W\right)}
f(W) \ .\label{slowpistonlimit}
\end{equation}
The factor $f(W)$ behaves like a rapidly oscillating function in this limit
(albeit the period increases as $n$ increases). 
We shall replace $f(W)$ by
its average value, given by the ratio of the area of a trapezoid in figure \ref{fig:overlapplot} to the
length of the base of the trapezoid,
$\frac{1+\alpha}{2+\alpha}$,
when we evaluate the integrals below.

One can derive the $W$ dependence of this distribution using adiabatic invariants,
except without the factor $f(W)$, as follows.

In \cite{Jar2}, Jarzynski considers as an example a single particle bouncing around
inside a three-dimensional cavity with hard walls, 
where the shape of the cavity is a function of a control parameter $\lambda$.
In it, he mentions the quantity $H_\lambda^{3/2}V_\lambda$ as an adiabatic invariant
(i.e. a constant when the control parameter is varied very slowly),
where $H_\lambda$ is the particle energy and $V_\lambda$ is the volume of the cavity.
These give the work performed on the particle with initial energy $E_0$:
$W_\infty=H_1-E_0=[(V_0/V_1)^{2/3}-1]E_0$
,
and the work distribution when $\lambda$ is switched infinitely slowly from $0$ to $1$:
\begin{equation}
\lim_{t_s\rightarrow\infty}\rho(W,t_s)=\left(\frac{4\beta^3W}{\pi r^3}\right)^{1/2}\exp{\left(-\frac{\beta W}{r}\right)}\theta(W)
\end{equation}
where $r=(V_0/V_1)^{2/3}-1$, $\beta=1/k_BT$, and $V_0>V_1$ (the cavity is being compressed).

We would like to employ the same idea here for the one-dimensional particle.

In one dimension, we take $H_\lambda^{1/2}L_\lambda$ as the invariant. This quantity is just the action or phase space area\cite{LL},
$\oint p\ dq/2\pi$, where $p$ and $q$ are momentum and position coordinates,
respectively.

The work done by the particle as the piston extends from $L$ to $(1+\alpha)L$, when the initial energy is $E_0$, is then
\begin{equation}
W=\left(1-\frac{1}{(1+\alpha)^2}\right)E_0
\end{equation}
The initial energy $E_0$ is boltzmann distributed. Multiplying the boltzmann factor by the density of states for a particle in one-dimension, $dp/dE\sim 1/\sqrt{E}$, produces the following expression
(apart from normalization factors):
\begin{equation}
\lim_{t_s\rightarrow\infty}\rho(W,t_s)\sim \frac{1}{\sqrt{W}}\exp{\left(-\beta W\frac{(1+\alpha)^2}{\alpha(2+\alpha)}\right)}
\end{equation}
which has the same functional dependence on $W$ as (\ref{slowpistonlimit})
when $\beta=1$, apart from the
overlap factor $f(W)$.

With the distribution (\ref{slowpistonlimit}), the
Jarzynski relation is verified explicitly as follows.
Using the integral
\begin{equation}
\int_0^\infty t^n e^{-st}dt=\frac{\Gamma(n+1)}{s^{n+1}}
\end{equation}
one obtains
\begin{eqnarray}
\left< e^W \right>&=&\int_0^\infty  P(W)e^WdW\nonumber\\
&=&\frac{\sqrt{\alpha(2+\alpha)/\pi}}{\alpha} \times \frac{1+\alpha}{2+\alpha} \times \frac{\Gamma(1/2)}{\frac{1}{[\alpha(2+\alpha)]^{1/2}}}\\
&=&1+\alpha = e^{-\Delta F}
\end{eqnarray}
where $\Delta F=-\ln{(1+\alpha)}=-\ln{(1+\Delta L/L)}$.


\end{document}